\documentclass[useAMS,usenatbib]{mn2e}
\usepackage{epsfig}
\usepackage{graphicx}
\usepackage{times}
\usepackage{color}

\def\cm3{cm$^{-3}$}
\def\kms{km~s$^{-1}$}

\def\lsun{L$_{\odot}$}
\def\rsun{R$_{\odot}$}

\def\msun{M$_{\odot}$}

\def\one{{\,\sc i}}
\def\two{{\,\sc ii}}
\def\three{{\,\sc iii}}

\def\beq{\begin{equation}}
\def\eeq{\end{equation}}

\def\zsun{Z$_{\odot}$}
\def\lesssim{\mathrel{\hbox{\rlap{\hbox{\lower4pt\hbox{$\sim$}}}\hbox{$<$}}}}
\def\gtrsim{\mathrel{\hbox{\rlap{\hbox{\lower4pt\hbox{$\sim$}}}\hbox{$>$}}}}

\def\apj{ApJ}
\def\apjs{ApJS}
\def\apjl{ApJL}
\def\aap{A\&A}
\def\araa{ARA\&A}

\def\mnras{MNRAS}
\def\nat{Nature}

\def\gray{$\gamma$-ray}
\def\grays{$\gamma$-rays}
\def\isoni{$^{56}{\rm Ni}$}
\def\isoco{$^{56}{\rm Co}$}

\title[Super-luminous supernovae]{Super-luminous supernovae: \isoni\ power versus magnetar radiation}
\author[Luc Dessart et al.]{\vspace{0.3cm} Luc Dessart,$^{1,2}$\thanks{email: Luc.Dessart@oamp.fr}
D. John Hillier,$^{3}$ Roni Waldman,$^{4}$ Eli Livne,$^4$ St\'ephane Blondin$^{5}$ \\
$^1$: Laboratoire d'Astrophysique de Marseille, Universit\'e Aix-Marseille \& CNRS,
UMR\,7326, 38 rue Fr\'ed\'eric Joliot-Curie, 13388 Marseille, France \\
$^2$: TAPIR, Mail code 350-17, California Institute of Technology, Pasadena, CA 91125, USA \\
$^3$: Department of Physics and Astronomy, University of Pittsburgh, USA  \\
$^4$: Racah Institute of Physics, The Hebrew University, Jerusalem, Israel \\
$^5$: Centre de Physique des Particules de Marseille, Universit\'e Aix-Marseille, CNRS/IN2P3, 163 avenue de Luminy,
13288 Marseille, France}

\date{Accepted 2012 August 05. Received 2012 July 02; in original form 2012 May 20}

\pagerange{\pageref{firstpage}--\pageref{lastpage}} \pubyear{2012}

\begin{document}

\maketitle

\label{firstpage}

\begin{abstract}
Much uncertainty surrounds the origin of super-luminous supernovae (SNe).
Motivated by the discovery of the Type Ic SN\,2007bi, we study
its  proposed association with a pair-instability SN (PISN). We compute stellar-evolution
models for primordial $\sim$\,200\,\msun\ stars, simulating the implosion/explosion
due to the pair-production instability, and use them as inputs for detailed non-LTE
time-dependent radiative-transfer simulations that include non-local energy deposition and
non-thermal processes.
We retrieve the basic morphology of PISN light curves from red-supergiant (RSG),
blue-supergiant (BSG), and Wolf-Rayet (WR) star progenitors.
Although we confirm that a progenitor 100\,\msun\ helium core (PISN model He100) fits
well the SN\,2007bi light curve, the low ratios of its kinetic energy
and \isoni\ mass to the ejecta mass, similar to standard core-collapse SNe, conspire to produce cool photospheres,
red spectra subject to strong line blanketing, and narrow line profiles, all conflicting with SN\,2007bi observations.
He-core models of increasing  \isoni-to-ejecta mass ratio have bluer spectra, but still too red to match SN\,2007bi,
even for model He125 -- the effect of \isoni\ heating is offset by the associated increase in blanketing.
In contrast, the delayed injection of energy by a magnetar represents a more attractive alternative to
reproduce the blue, weakly-blanketed, and broad-lined spectra of super-luminous SNe.
The extra heat source is free of blanketing and is not explicitly tied to the ejecta.
Experimenting with a $\sim$\,9\,\msun\ WR-star progenitor, initially exploded
to yield a $\sim$1.6\,B SN Ib/c ejecta but later influenced by tunable magnetar-like radiation,
we produce a diversity of blue spectral morphologies  reminiscent of SN\,2007bi, the peculiar
Type Ib SN\,2005bf, and super-luminous SN\,2005ap-like events.
\end{abstract}


\section{Introduction}

In the last decade, a number of super-luminous supernovae (SNe) have been identified but their origin remains
highly uncertain. Most of these exhibit the Type Ia/Ib/Ic SN light-curve morphology, merely ``expanded''
to form a broader and brighter peak.
In this category, we find a zoo of events including, e.g., the Type IIn SN\,2006gy \citep{smith_etal_07a},
the peculiar type Ib SN\,2005bf \citep{folatelli_etal_06,maeda_etal_07} with its double-peak light curve, the type Ic SN\,2007bi
\citep{galyam_etal_09} with its extended nebular tail, the linearly-declining SN\,2008es \citep{gezari_etal_09},
as well as a rather uniform group of SNe at redshifts of 0.2--1.2 with a unique, nearly-featureless,  blue
continuum, and a fast fading nebular flux \citep{quimby_etal_11}.
Today, three mechanisms are proposed for these super-luminous events, representing
extreme versions of SNe interacting with a circumstellar medium, \isoni-powered SNe, and magnetar-powered SNe.
For brevity, and since an interaction mechanism is believed not to be relevant to SN\,2007bi, we restrict our discussion
to the last two mechanisms.

 Historically, the most natural way to explain a large SN luminosity is the production of a larger-than-average
 \isoni\ mass \citep{colgate_mckee_69}. The essential feature is that because the half-life of \isoni\ is 6.075\,d
 and that of its daughter nucleus \isoco\ is 77.23\,d, radioactive-decay energy can reheat the ejecta once it has expanded
 to $\gtrsim$10$^{14}$\,cm and become less sensitive to $PdV$ losses.
 However, ``standard" core-collapse SNe are generally ineffective \isoni\ producers --
the \isoni\ production is strongly conditioned and limited by the explosion energy and the progenitor structure.
  Today, highly energetic magneto-rotational explosions are expected to be the most suitable means to produce
  super-luminous SNe \citep{burrows_etal_07b}. These are generally nicknamed ``hypernovae",
   and are distinct from the more germane neutrino-driven core-collapse SN explosions \citep{buras:06b}.

    Pair-instability SNe (PISNe) represent an alternative for producing a large amount of \isoni.
   In the exceptional instance of stars with a main-sequence mass in the range 140--260\,\msun, expected
  to form at low metallicity in the early Universe  \citep{bromm_larson_04},
  e$^{-}$e$^{+}$ pair production may lead to an explosion and give rise to a PISN
  \citep{barkat_etal_67,HW02,langer_etal_07,waldman_08}.
  Although the explosion mechanism is robust, it is still unclear whether such massive stars can form.
  If they do form, their mass loss is of concern as it can considerably affect the final
  stellar mass and radius, and thus, the resulting explosion properties, SN radiation, and detectability
  \citep{scannapieco_etal_05,kasen_etal_11}.

 An alternative means to produce a bright display is by magnetar radiation \citep{wheeler_etal_00,maeda_etal_07,
 kasen_bildsten_10,woosley_10}. The energy lost in the process leads
 to the spin down of the magnetar, which eventually quenches its power.
 For a dipole field, the spin-down time scale is $t_{\rm sp}\sim$\,4.8$B^{-2}_{15}P^2_{10}$\,d, where $B_{\rm 15}$ is
 the magnetic-field strength in 10$^{15}$\,G and $P_{\rm 10}$ the rotation period $P$ in units of 10\,ms.
 For suitable choices of $B_{15}$ and $P_{\rm 10}$, this timescale can be comparable to the half-life
 of \isoni/\isoco\ and consequently makes magnetar radiation an attractive substitute for long-lived super-luminous
 SNe - in combination with different ejecta masses, it also provides a natural modulation for the time to peak brightness,
 the luminosity at peak \citep{kasen_bildsten_10}, as well as for the fading rate from peak.

\begin{figure}
\epsfig{file=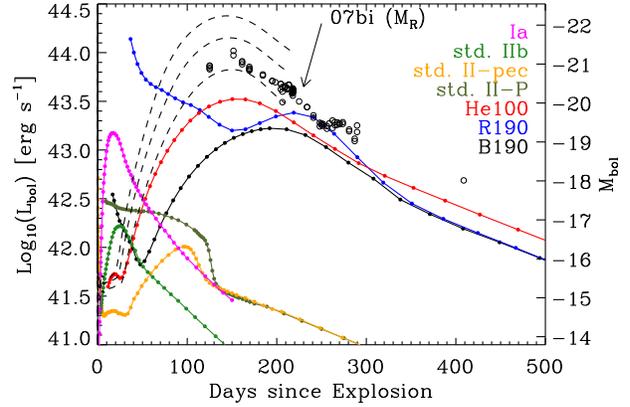,width=8.5cm,bbllx=0,bblly=0,bburx=510,bbury=340,clip=}
\caption{Synthetic bolometric light curves extracted from our {\sc cmfgen} simulations
of PISNe and other ``standard'' SN models (solid). We include {\sc v1d}  light curves for models He105, He115,
and He125 of increasing peak luminosity (dashed) and the estimated SN\,2007bi absolute $R$-band
magnitude \citep[circles]{galyam_etal_09}.
\label{fig_lbol}
 }
\end{figure}

The proposition that SN\,2007bi is a PISN is controversial.
It was discovered around the peak of the light curve ($M_R\sim-$\,21\,mag), revealed a
slowly fading $R$-band magnitude consistent with full \gray\ trapping from
$\sim$\,5\,\msun\ of \isoni\ (Fig.~\ref{fig_lbol}).
It exploded in an environment with a metallicity of one third solar,  which conflicts with
 star formation \citep{bromm_larson_04} and evolution theory \citep{langer_etal_07},
 which expect such stars to form and explode as a Type Ic SN at much lower metallicity only.
\citet{galyam_etal_09} performed a few simulations for SN\,2007bi, covering a range of progenitor He cores and thus
explosion characteristics, and found their He100 model to be adequate. From their modeling of a nebular-phase
spectrum, they infer ejecta masses compatible with the PISN scenario.
Improving upon the original work of \citet{scannapieco_etal_05}, \citet{kasen_etal_11} studied
a broad mass range of PISN progenitor models including RSG, BSG, and WR stars.
Their He100 model gives a suitable match to the SN\,2007bi light curve, as well as a rough agreement with the near-peak spectrum.

An alternative scenario, involving the collapse of a massive-star core, has been proposed by \citet{moriya_etal_10}.
In association with the extreme properties of the SN explosion they also invoke radioactive-decay energy
to explain the light curve.
The situation remains blurred, epitomized by the rough compatibility of both the \isoni\ model (PISN model He100 or
extreme core-collapse SN)  and the magnetar model for explaining the SN\,2007bi light curve
\citep{kasen_bildsten_10,kasen_etal_11}. Understanding what distinguishes these different scenarios
is thus critical to identify the nature of super-luminous SNe.

In the next section we present the numerical setup for the in-depth study of PISN explosions that we have
undertaken, and which will be discussed more fully elsewhere (Dessart et al., in prep).
Simulation results for three different PISN progenitor models are presented in Sect.~\ref{sect_results}.
We then focus on the model  He100 that was proposed for SN\,2007bi, and discuss its incompatibilities
with the observations. Ways of alleviating these incompatibilities are discussed in Sect.~\ref{sect_slsn}.
In particular we propose two means to produce a super-luminous SN with a bluer color -- either through
an increase in the \isoni-mass to ejecta mass ratio or, alternatively, through a delayed energy injection
from the compact remnant (Sect.~\ref{sect_slsn}).

\section{Numerical setup of PISN simulations}

   The work presented in this paper was produced in several independent steps. First, a large grid of massive-star progenitors
   with main-sequence masses in the range 160--230\,\msun\ were evolved with MESA \citep{paxton_etal_11} until central
   $^{20}$Ne exhaustion,  assuming 10$^{-4}$\,\zsun\, and no rotation.
   At that time, these stars are either RSG, BSG, or WR stars.
   These simulations are then remapped onto the 1D radiation-hydrodynamical code {\sc v1d} \citep{livne_93},
   with allowance for (explosive) nuclear burning and radiation transport.
   At $^{20}$Ne exhaustion, oxygen is naturally ignited and consumed in less than a minute to \isoni\ and intermediate-mass
   elements (IMEs). We model this explosion phase and the later evolution of the ejecta for a few years (Waldman et al., in prep.).

   Here, we focus on three PISN types:
   1) II-P: Model R190 from a 190\,\msun\ main-sequence star and dying as a 164.1\,\msun\ RSG with a surface radius
   of 4044\,\rsun;
   2) II-pec: Model B190 from a 190\,\msun\ main-sequence star and dying as a 133.9\,\msun\ BSG with a surface radius
   of 186.1\,\rsun;
   3) Ic (or Ib): Model He100 from a 100\,\msun\ helium core  (190\,\msun\ initially, artificially stripped of its
   hydrogen envelope, but without subsequent mass loss) and dying as a 100\,\msun\ WR with a surface radius of 1.2\,\rsun.
   In the same order, these models synthesize 2.63, 2.99, and 5.02\,\msun\ of \isoni, have  kinetic energies of
   33.2, 34.5, and 37.6\,B, and a representative expansion velocity $v_{\rm rep}  \equiv \sqrt{2E/M}  \sim 5000$\,\kms.

   Although PISNe are thermonuclear explosions, akin to SNe Ia,
   burning only occurs in the inner ejecta so that \isoni\ is ultimately confined to shells moving with velocities inferior
   to 2000\,\kms\ (models R190 and B190) and 4000\,\kms\ (He100).
   In the second part of this letter, we present additional He models, similar to He100, but with initial masses
   from 105 to 125\,\msun, spaced every 5\,\msun\ (Sect.~\ref{sect_slsn}).
   The properties of our hydrodynamical input models are comparable to those of \citet{kasen_etal_11}, so we defer
   their detailed presentation to Waldman et al. (in prep.).

   For the radiative-transfer calculations of light curves and spectra, we adopt our standard procedure
   \citep{DH10,DH11,dessart_etal_11,dessart_etal_12}.
   \citet{HD12} have recently given a full description of the code {\sc cmfgen} for SN calculations.
We perform time-dependent simulations for the full ejecta, from the photospheric to the nebular phase.
We start when homologous expansion is reached (a few days to a few weeks after explosion) and use as initial
conditions the ejecta chemical stratification and structure computed with {\sc v1d}.
  Our radiative-transfer simulations are fully time-dependent, non-LTE, include non-local energy deposition,
  treat non-thermal processes associated with \grays, and yield synthetic spectra from the far-UV to the far-IR.
  We treat explicitly the effects of line blanketing arising from the presence of a multitude of optically-thick
  lines of metals. Expansion strengthens line blanketing by broadening the effective width of lines, increasing
  the occurrence of line overlap \citep{HM98_lb}.

 \section{Results and comparison to SN\,2007\lowercase{bi}}
\label{sect_results}

   The synthetic light curves computed for PISN models R190, B190, and He100 reflect their large ejecta and
   \isoni\ masses, and are characterized by a high-brightness phase that lasts a few hundred days. Their peak
   luminosities are on the order of 10$^{10}$\,\lsun. They transition  to the nebular  phase, with a luminosity
   reflecting the rate of \isoco\ decay-energy release, $\sim$\,200\,d after peak (Fig.~\ref{fig_lbol}).
   The morphological diversity of PISN light curves is analogous to that of SNe observed in the local Universe,
    belonging to the Type II-P class (R190),  to the Type II-pec class (B190),  and to the Type Ib/c class (He100 and analogs), in
    qualitative agreement with the RSG, BSG, and WR star progenitor, and in quantitative agreement with the
    simulations of \citet{kasen_etal_11}. For comparison we add the lower-energy lower-mass counterparts
    for each SN type \citep{DH10,DH11,dessart_etal_11}, as well as our results for a delayed-detonation model
    of a Chandrasekhar-mass white dwarf with 0.67\,\msun\ of \isoni\  (Dessart et al., in prep).

      These PISNe reach photospheric
   radii of $\sim$\,10$^{16}$\,cm, but apart from the R190 model which retains a hot photosphere for many weeks after the explosion,
   we find that PISN photospheres are typically cold. On the rise to peak brightness, the photospheres of models B190 and He100 heat up
   from $\sim$\,4000\,K to $\sim$6000\,K. Fading after peak, all PISN photospheres cool to $\sim$4000\,K within
   100-200\,d. As they turn nebular, there is no longer any photosphere, nor any trapped energy, and the ejecta continue
   to cool, reaching typically $\gtrsim$\,1000\,K three years after the explosion.
   So, the large initial \isoni\ mass that causes the large peak brightness is no guarantee
   of high photospheric temperatures.

\begin{figure}
\epsfig{file=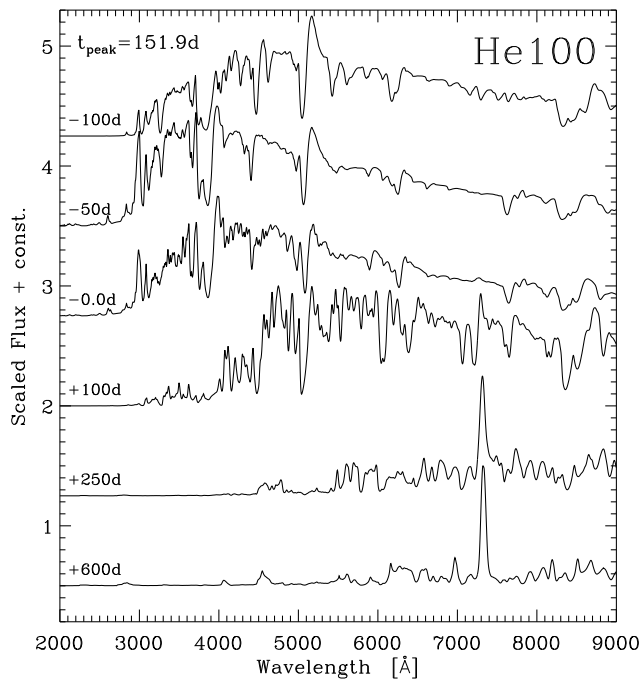,width=8.5cm,bbllx=45,bblly=35,bburx=566,bbury=613,clip=}
\caption{Montage of synthetic spectra for model He100. Labels on the left give the days since the bolometric maximum
at 151.9\,d after explosion.
\label{fig_spec}
 }
\end{figure}

  Spectroscopically, our PISN simulations reflect in many ways the properties
  of lower-mass lower-energy SNe II-P, II-pec, and Ib/c \citep{dessart_etal_12}.
  For model He100, the proposed model for SN\,2007bi,
  our synthetic spectra become bluer on the rise to peak due to decay heating (Fig.~\ref{fig_spec}).
  Bound-bound and bound-free processes in IMEs dominate the opacity,
  producing lines of Ca\two, Mg\one, Mg\two, Si\two, and O\one.
  As the photosphere recedes deeper into the ejecta, at and beyond the peak of the light curve, the mass fraction
  of  iron-group elements (IGEs) in the photosphere increases, and so do the effects of line blanketing, first by Fe\two,
  and later by Fe\one.
  Exacerbated by the photosphere cooling, the emergent spectrum reddens, fading strongly short-ward of 5000\,\AA.
  By 500\,d after explosion, the spectrum is nebular and reveals lines of O\one\ and Ca\two.

   With a peak $M_{\rm bol}\sim-$20.2 ($M_R=-$20.5) and slow nebular fading, our model He100 reproduces
   well the light curve morphology of SN\,2007bi, although it is too faint by $\sim$\,0.8\,mag in $R$.
   The red color of model He100 at 54\,d after peak ($B-R=$\,1.47) strongly contrasts with the blue color of
   SN\,2007bi at that time (we estimate $B-R\sim$\,0.45\,mag from the observed spectrum while
   \citealt{Y10_full} infer a K-corrected $B-R$ of 0.23\,mag).

  Further beyond the peak, model He100 shows a spectrum strongly blanketed by Fe\two\ and Fe\one, with narrow lines
  that form in ejecta regions moving at $\lesssim$\,4000\,\kms, and little flux shortward of 5000\,\AA\ (Fig.~\ref{fig_spec}).
  This contrasts with SN\,2007bi which shows a sustained blue spectrum with significant flux shortward
  of 3500\,\AA\ and broad lines of Ca\two, Fe\two, Si\two,
  which form in an ejecta expanding at $\gtrsim$\,6000\,\kms\ \citep{galyam_etal_09}.
  Model He100, proposed by \citet{galyam_etal_09} as viable for SN\,2007bi, is thus disfavored on numerous grounds.
  While \isoni\ is a suitable energy source for producing a super-luminous SN,  the associated IGE line blanketing and cool
  photospheric temperatures after the light-curve peak conspire to produce spectra that are red, rather
  than blue as observed in SN\,2007bi, or generally in super-luminous SNe.

    The apparently satisfactory spectral fit obtained by \citet{kasen_etal_11} results from the inadequate choice of epoch
    for their He100 model spectrum, as they use a 50-d pre-peak synthetic spectrum to match the 50-d post-peak spectrum of SN\,2007bi.
    The  color change as the PISN model bridges the peak was thus unfortunately ignored.
   \citet{galyam_etal_09} find a match to the low-quality nebular-phase spectrum but do not model the earlier
   higher-quality observations nor do they explain the early blue colors, the weak blanketing, the broad features,
   or the apparent transition to a nebular spectrum between $\sim$400-500\,d after peak.
   Their estimate of a 100\,B kinetic energy (ignoring the $\sim$\,10\,B binding energy of the progenitor)
   requires a much more efficient burning than they claim, yielding not  $\sim$5\,\msun\ as suggested but
   few 10\,\msun\ of \isoni\ --- this then places the luminosity at odds with the SN\,2007bi light curve.

   Proposing a consistent PISN model
   that matches the light curve, the color evolution, and the spectral properties (ions, line-profile widths and strengths etc.)
   as well as agrees with massive-star evolution and formation is a considerable challenge.
   As a reminder, standard stellar-evolutionary models do not support the existence of Type Ic PISNe at metallicities
   as high as \zsun/3 \citep{langer_etal_07}, even if suitable super-massive progenitors existed.
   In the next section,
   we discuss two different models that may cure some of the discrepancies  between model He100
   and SN\,2007bi, and also generalize our ideas to other, non-interacting, super-luminous SNe.

  \begin{figure}
\epsfig{file=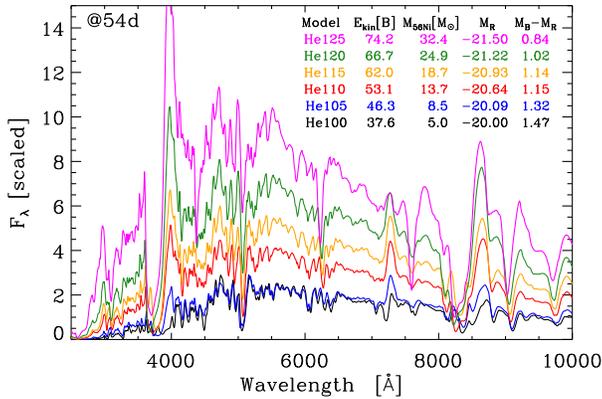,width=8.cm,bbllx=20,bblly=0,bburx=510,bbury=340,clip=}
\caption{
Montage of synthetic spectra for models He100 to He125 at 54\,d after peak.
Models of increasing $M_{^{56}\rm Ni}/M_{\rm ejecta}$ have bluer
colors, although redder than SN\,2007bi, whose $B-R \sim$\,0.45 at that time.
\label{fig_spec_all}
 }
\end{figure}

\section{Blue-colored super-luminous SN\lowercase{e}}
\label{sect_slsn}

    Despite the large initial \isoni\ mass, the large ejecta mass in model He100 renders the heating from decay
    ineffective at producing and sustaining a hot photosphere/ejecta.
    One remedy is to increase the \isoni\ mass to ejecta mass ratio, from the 5\% in model He100,
    quite typical of core-collapse SNe, to few tens of percent, more typical of SNe Ia.
    In SNe Ia, blanketing is certainly severe, but the enhanced heating
    produces relatively bluer colors. This in part stems from the lower mass of the ejecta, which turns
    transparent at $\sim$30\,d (rather than $\sim$300\,d), i.e., when the decay-energy rate is $\sim$\,10 times larger.

    To experiment with this idea, we ran models similar to He100 but with masses between 105 and 125\,\msun.
    When exploding as PISNe, these synthesize between 8.5 and 32.4\,\msun\ of \isoni.
    The spectra at 54\,d post peak (selected to correspond to the first spectrum of 07bi) are now bluer, and the more
    so as the \isoni-to-ejecta mass ratio increases from 8 to 26\% (Fig.~\ref{fig_spec_all}; labels give additional model properties).
    While models He105-He115 better match the luminosity of SN\,2007bi (Fig.~\ref{fig_lbol}), they retain a similar, and discrepant,
    color as model He100. For model He125, despite obvious signs of blanketing by Fe\two, the color
    matches more closely that of SN\,2007bi, but this model is now 1\,mag too bright in the $R$-band.
    In a similar fashion to model He100 and SNe Ia, it will not only be heated by 32\,\msun\ of \isoni/\isoco, but it will
    also be drastically reddened by 32\,\msun\ of iron $\gtrsim$\,200\,d after explosion.

   What emerges, though, is the difficulty, even for extreme \isoni\ yields, to produce
   high photospheric temperatures. Being fundamentally associated with large ejecta masses and huge explosion
   energies, PISN ejecta inevitably recombine: their
   photospheres adjust to reside in the transition zone between optically thick and thin conditions.
   Consequently, their spectra tend to peak in the optical, reflecting the low recombination
   temperature of the dominant ion at the photosphere (as in SNe II-P; \citealt{DH11}).

   This difficulty of the PISN scenario motivated us to explore the radiative signatures that would result from
   delayed energy injection, as in a magnetar.
   \citet{woosley_10} studied this context in otherwise ``standard" core-collapse SN explosions, although
   he focused exclusively on the resulting bolometric light curve.
   We already know that with arbitrary choices of ejecta mass,
   magnetic field, and initial spin, a broad diversity of luminous light curves can be produced \citep{kasen_bildsten_10}.
   Here, we investigate the spectroscopic properties of an ejecta subject to
   an energy injection $E_{\rm dep}$ at a constant rate up to a post-explosion time $\delta t$ (given the many
   parameters tuning this scenario, it is not critical for our exploration to be more accurate). We simulate this scenario with {\sc v1d} using
   the approach of \citet{DLW10b} starting from a 1.56\,B 6.94\,\msun\ ejecta (with 0.17\,\msun\ of \isoni) produced
   from the explosion of an 8.74\,\msun\ WC star (model s40 of \citealt{WHW02}).
   Using a unique $\delta t=$\,20\,d, we compute models with $E_{\rm dep}=$\,0.1, 0.3, and 1.0\,B
   (models pm0p1, pm0p3, and pm1p0), producing 40-d broad bolometric light-curve peaks $\sim$30\,d after explosion
   with representative luminosities of 10$^9$ to 10$^{10}$\,\lsun.
   Were these luminosities powered by \isoni\ decay alone, they would require $\gtrsim$\,1\,\msun\ of \isoni, an amount
   anomalously large for standard Type Ib/c SNe.

\begin{figure}
\epsfig{file=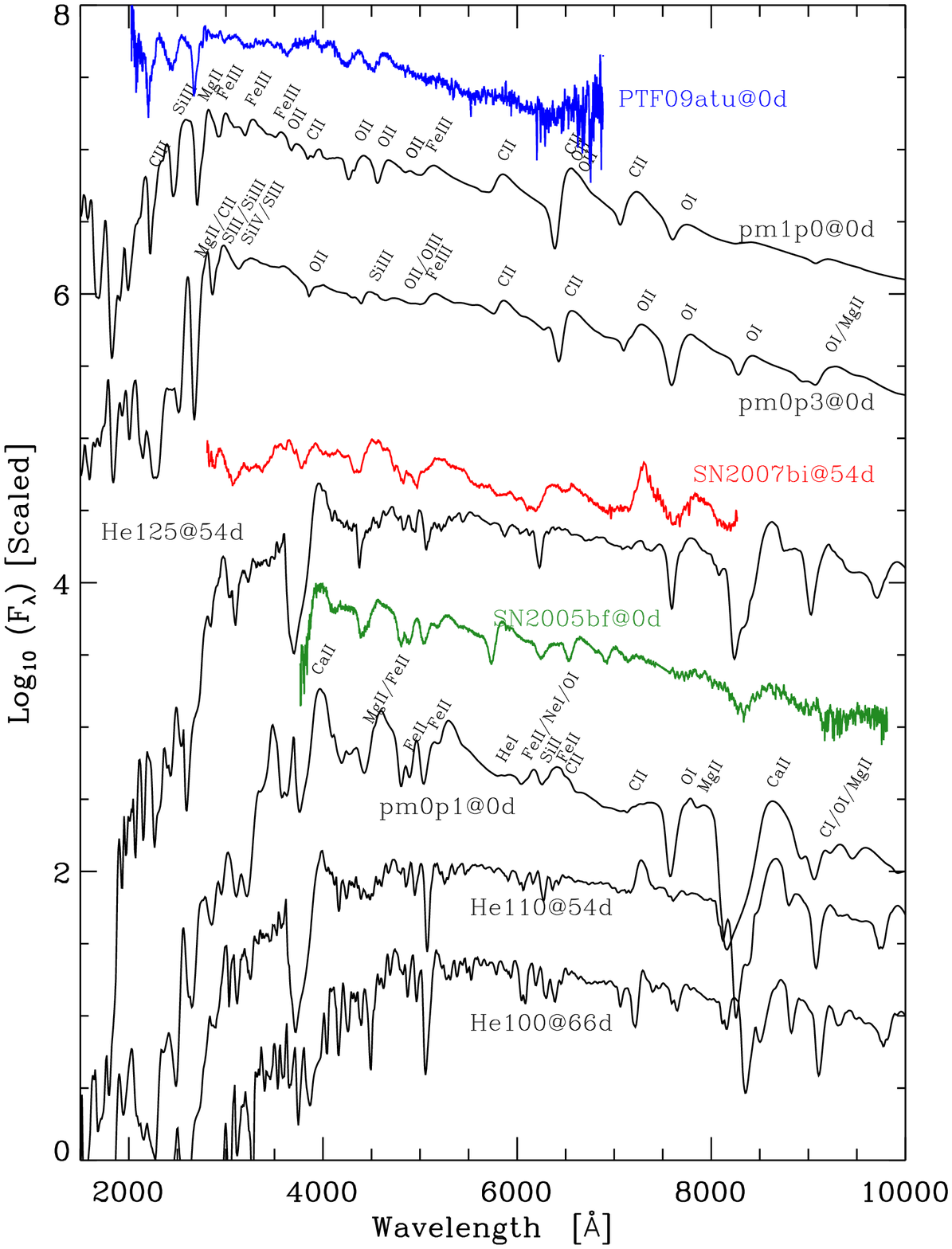,width=8.5cm,bbllx=20,bblly=70,bburx=586,bbury=763,clip=}
\caption{Comparison between SNe PTF\,09atu, 2007bi, and 2005bf, synthetic spectra of
PISN models He100, He110, and He125, and magnetar-powered models pm0p1, pm0p3, and pm1p0.
Some dust extinction ($E(B-V)=$\,0.4\,mag) is applied to the last two models for convenience.
Labels indicate the time since bolometric maximum.
\label{fig_pm}
}
\end{figure}

    Using the {\sc v1d} ejecta structure around the broad light-curve peak,  we generate steady-state non-LTE spectra
    with {\sc cmfgen} (Fig.~\ref{fig_pm}).
    For the lowest energy injection (model pm0p1), the heating is very weak and produces hardly any hardening of the
    spectrum compared to the default model with no energy injection (thus not shown).
    However, by increasing the deposition of energy from 0.1\,B (pm0p1) to 1.0\,B (pm1p0), the colors of the resulting
    spectrum dramatically harden. We obtain bluer spectra roughly compatible with SN\,2007bi, the peculiar SN Ib
    2005bf \citep{folatelli_etal_06}, and even the 2005ap-like event PTF\,09atu \citep{quimby_etal_11}.
    These three models mimic the effect of magnetars/pulsars of varying powers, something not unexpected from nature.

    SNe that are super luminous due to the magnetar-energy deposition do not suffer from
    excess line blanketing inherent to an anomalously large production of \isoni, which is inevitable in PISNe and
    other extreme core-collapse SN events. Instead, the extra energy injected heats the material and thermally
    excites the gas (provided thermalization takes place), producing lines of Si\two, C\two, and He\one\
    (SNe\, 2005bf and 2007bi; model pm0p1), and in extreme cases O\two, C\two, Si\three, and Fe\three\ (PTF\,09atu; model pm1p0).
    This exploration is not fully satisfactory (e.g., [Ca\two]\,7300\,\AA\ is not predicted in pm0p1 but is seen in SN\,2007bi),
    but it is indicative --- the spectra of these models are more fundamentally in agreement with SN\,2007bi than
    \isoni-powered super-luminous SN models.
    The 0.11\,\msun\ of helium in our ``pm'' models is thermally excited and produces He\one\ lines.
    The temperatures are, however, too low, even in model pm1p0, to produce He\two\ emission (e.g., at 4686\,\AA),
    as observed in SN\,2008es \citep{gezari_etal_09}.

    In contrast to the irrevocably high-mass PISNe, magnetar-powered ejecta, which can be of any mass,
    can naturally produce broad lines at all times. The large energy needed
    to power the light curve launches a snow-plow of the inner ejecta layers, forming a dense shell at constant velocity.
    This inner shell velocity is 3000\,\kms\ in model pm0p1, 3800\,\kms\ in pm0p3, and 5000\,\kms\ in pm1p0.
    In SN\,2007bi, the width of spectral features hardly changes from 54\,d until 414\,d after the light-curve peak, something
    difficult to explain with the He100 PISN model (Fig.~\ref{fig_spec}).
    Non-local energy deposition can counteract the recession of the spectral formation region, but this
    influences the most optically-thick lines and will struggle to produce broad nebular lines in such massive
    PISN ejecta; its impact is inhibited due to the low-metallicity of our PISN model.

    An interesting issue about super-luminous SNe is their systematic detection near, or after, the peak
    of the light curve. In the \isoni-powered model, the heat generated at depth in these massive ejecta conspires
    to produce a near symmetric light-curve peak (Fig.~\ref{fig_lbol}), and so one would expect to discover a
    fraction of these on the relatively long rise to peak. In the magnetar model, the lack of a pre-peak detection
    is naturally explained: The relatively large $B$ and $\Omega$ needed to power the light curve imply a fast spin down,
    and a fast rise to the light-curve peak is compatible with a low/moderate mass ejecta.
    Similarly, the post-peak fading may  cover a range of slopes reflecting the
    differing instantaneous contributions of \isoni\ and magnetar-energy injection.

    The magnetar model is also well supported by the large number of such objects in the Galaxy \citep{muno_etal_08}.
     They are obviously easy to form, in contrast with PISNe, expected to exist primarily in the Early
    Universe.
     They are also routinely produced at low metallicity by massive-star evolution with fast rotation \citep{woosley:06,georgy_etal_09},
    perhaps providing an alternate channel to black-hole formation \citep{dessart_etal_08a,metzger_etal_11,dessart_etal_12b}.

     Future work requires the modeling of magnetar radiation in various massive-star progenitors using {\sc v1d}
   and {\sc cmfgen} to characterize the range of super-luminous SNe  this  scenario can produce in terms of ejecta,
   spectral, and light-curve properties, e.g., rise time, peak luminosity, color, and fading-rate at nebular times.

\section*{Acknowledgments}

LD and SB acknowledge support from European-Community grant PIRG04-GA-2008-239184
and from ANR grant  2011-Blanc-SIMI-5-6-007-01.
DJH acknowledges support from STScI theory grants HST-AR-11756.01.A and  HST-AR-12640.01,
and NASA theory grant NNX10AC80G.

\label{lastpage}

\end{document}